\documentclass[prl,twocolumn,a4paper,showpacs]{revtex4-1}
\usepackage{amssymb}
\usepackage{wasysym}
\usepackage{graphicx}
\usepackage{epsfig}
\usepackage{subfigure}
\begin{document}

\title{The nature of the continuous nonequilibrium phase transition of Axelrod's model}

\author{Lucas R. Peres and Jos\'e F. Fontanari}
\affiliation{Instituto de F\'{\i}sica de S\~ao Carlos,
  Universidade de S\~ao Paulo,
  Caixa Postal 369, 13560-970 S\~ao Carlos, S\~ao Paulo, Brazil}


\begin{abstract}
Axelrod's model in the square lattice with nearest-neighbors interactions  exhibits 
culturally homogeneous  as well as culturally fragmented absorbing configurations. In the case the agents are characterized by $F=2$ cultural features and  each feature assumes 
 $k$  states drawn from a Poisson distribution of parameter $q$ these  regimes  are separated by a continuous  
transition at $q_c = 3.10 \pm 0.02$. Using Monte Carlo simulations and finite size scaling we show that the mean density of cultural domains $\mu$ is an order parameter of the model
that vanishes as $\mu \sim \left ( q - q_c \right)^\beta$  with $\beta = 0.67 \pm 0.01$ at the critical point.  In addition, for the correlation length critical exponent we find $\nu = 1.63  \pm 0.04$  and for Fisher's exponent, $\tau = 1.76 \pm 0.01$. This set of critical exponents places the continuous
phase transition of Axelrod's model apart from the known universality classes  of nonequilibrium lattice   models.

\end{abstract}

\maketitle

Social influence  and homophily (i.e., 
the tendency of individuals to interact preferentially with similar others)
have long  been  acknowledged as major factors that influence the persistence of cultural diversity  in a community
\cite{Lazarsfeld_48,Castellano_09}.
The  manner these factors affect diversity, however,  has begun to be understood quantitatively  after the proposal of 
an agent-based model  by the political scientist Robert Axelrod in the late 1990s only \cite{Axelrod_97}.
In Axelrod's   model,   the agents  are represented by  strings  of
cultural features of length $F$, where each feature can adopt a certain number $k$ of distinct states
(i.e., $k$ is  the common number of states each feature can assume). The term culture is
used to indicate the set of individual attributes that are susceptible to social influence. The homophily factor is  taken into account
by assuming that the interaction between  two agents takes place
with probability proportional to their cultural similarity  (i.e., proportional to the number of states they have in common), whereas social
influence 
enters the model by allowing the agents to become more similar when they interact. Hence there is a  positive feedback loop between homophily and social influence: similarity leads to interaction, and interaction leads to still more similarity. Overall, the
conclusion was that the homophilic interactions together with the limited range of the  agents' interactions favor multicultural steady
states  \cite{Axelrod_97}, whereas relaxation of these conditions favors  cultural homogenization \cite{Klemm_03a}.

In Axelrod's model, there are two types
of absorbing configurations in the thermodynamic limit \cite{Castellano_00,Klemm_03,Vazquez_07,Barbosa_09}: the ordered  configurations, which  are characterized by the presence of
few cultural domains of macroscopic size,   and the disordered absorbing configurations, where all domains are microscopic. In time, a cultural domain is defined  as  a bounded region of uniform culture. 
According to the rules of the model, two neighboring agents that do not have any cultural feature
in common are not allowed to interact and the interaction between agents who share all their cultural features produces no changes.
 Hence at the stationary state we can guarantee that any pair of neighbors are either identical 
or completely different  regarding their cultural features. In fact,
a feature that sets Axelrod's model  apart from most
lattice models that exhibit nonequilibrium phase transitions \cite{Marro_99} is  that all stationary 
states of the dynamics are absorbing states, i.e., the dynamics always freezes in one of these states. This
contrasts with lattice models that exhibit an active state in addition to infinitely many  absorbing states \cite{Jensen_93}
and the phase transition occurs between the active state and the (equivalent) absorbing states. 
In Axelrod's model, the competition between the disorder of the initial configuration that favors cultural fragmentation and  the ordering bias of social influence 
that favors homogenization results in the phase transition between those two classes  of absorbing states in the square lattice \cite{Castellano_00}.  Since the transition occurs in the properties of the absorbing states, it  is static in nature \cite{Vilone_02}.

Here we address a variant of Axelrod's model proposed by Castellano et al.\ that is more suitable for the study of the phase transition \cite{Castellano_00}. In the original
Axelrod's model, the initial states of the $F$  cultural  features of the  agents are  drawn randomly from a uniform distribution on the integers $1, 2, \ldots, \hat{q}$. Since both parameters of the model -- $\hat{q}$ and $F$ -- are integers,  it is not possible to determine whether the transition is continuous or not, let alone
to say something meaningful about its class of universality. A  way to circumvent this problem is to draw the  initial integer values (states) of the cultural  features using a Poisson distribution of parameter $q \in \left [ 0, \infty \right )$,
\begin{equation}\label{Poisson}
P_k = \exp \left ( -q \right ) \frac{q^k}{k!}
\end{equation}
with $k=0,1,2, \ldots$.  As in the case the states are chosen from a uniform distribution, Castellano et al.\ showed that the Poisson variant exhibits a phase transition in the square lattice with the bonus that they were also able to show that the transition  is continuous for $F=2$ and discontinuous for $F > 2$ \cite{Castellano_00}. 
 Here we focus on the continuous transition for $F=2$  in the square lattice of size $L \times L$ with periodic boundary conditions using extensive Monte Carlo simulations of lattices of linear size up to $ L= 1200$. We show that this transition takes place at $q= q_c = 3.10 \pm 0.02$ and determine the
critical exponents that characterize the model near the critical point.

The Poisson variant differs from the original Axelrod model only by the procedure used to generate the cultural 
states of the agents at the beginning of the simulation. Once the initial configuration is set, the dynamics  proceeds
as in the original model \cite{Axelrod_97}. In particular,   at each time we pick an agent at random
(this is the target agent) as well as one of its neighbors. These two 
agents interact with probability equal to  their cultural similarity, defined as the fraction of 
identical features in their cultural strings.   An interaction consists of selecting at random one of the distinct features, and making the
selected feature of the target agent equal  to  the corresponding  feature of its neighbor. This procedure is repeated until 
the system is frozen into an absorbing configuration. 

Once an absorbing state is reached we count the number of  cultural domains ($\mathcal{N}$) and record the size of the largest one
($\mathcal{S}_{max}$).  Average of these quantities over a large number of independent runs, which differ by the choice of the initial cultural states of the agents as well as by their  update sequence, yields the measures  we use to characterize the statistical properties of the absorbing configurations.

\begin{figure}[!ht]
  \begin{center}
\includegraphics[width=0.48\textwidth]{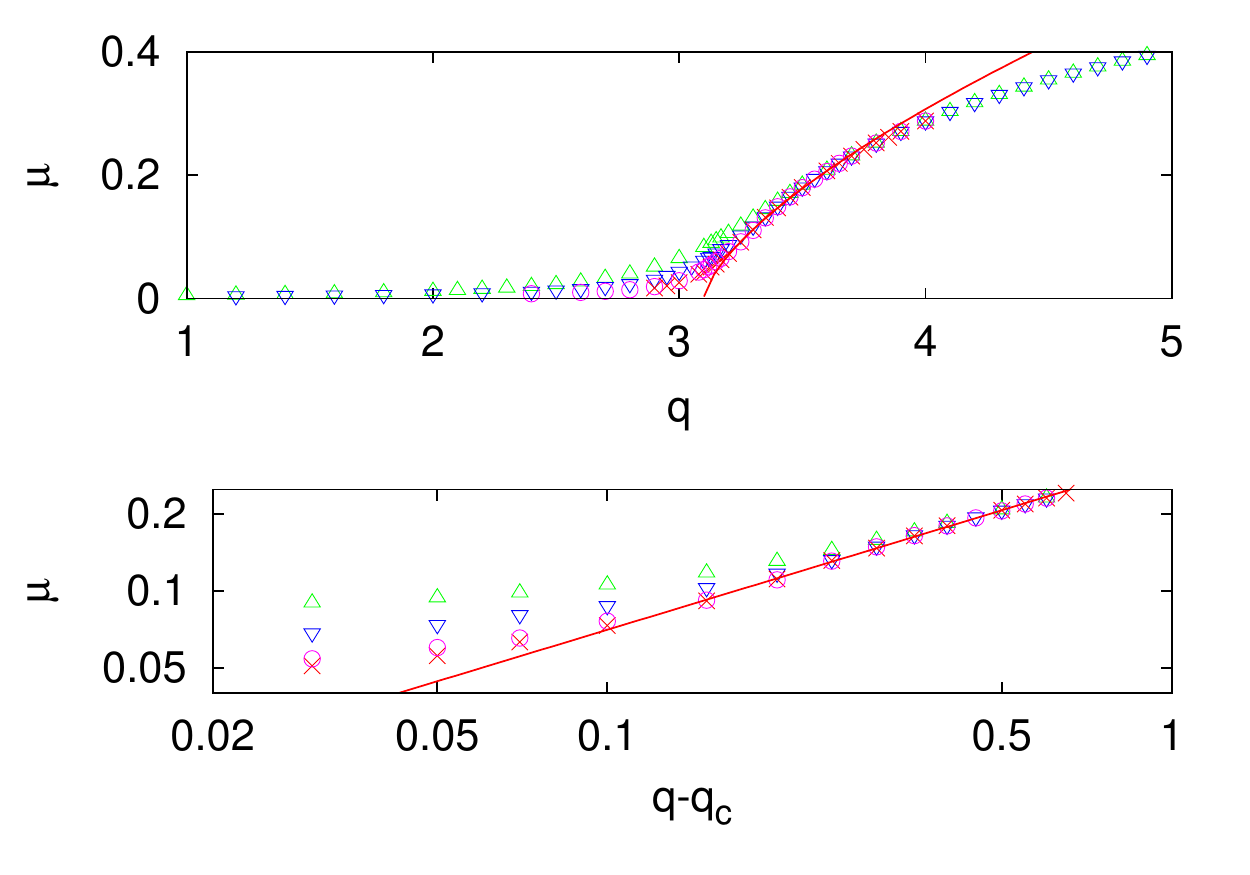}
  \end{center}
\caption{Upper panel: Mean density of domains $\mu$  as function of the Poisson parameter $q$
for  lattices of linear  size $L =  200 $ ($\triangle$), $L=400$ ($\triangledown$), $L=800$ ($\bigcirc$)  and $L=1000$ ($\times$).   
Lower panel: Log-log plot of $\mu$ against $q-q_c$ for $q_c= 3.1$.  
The solid line in both panels  is the two-parameters fitting  function $\mu= \mathcal{A} \left ( q - {q}_c \right )^{\beta}$ with $\mathcal{A}=0.331 \pm 0.003$,  and $\beta= 0.67 \pm 0.01$.
The error bars are smaller than the symbol sizes. }
\label{fig:1}
\end{figure}

Let us consider first  the mean density of domains   $\mu = \left \langle \mathcal{N} \right \rangle/L^2$.   This quantity is important
because it determines whether the number of domains is extensive or not in the thermodynamic limit. In the standard percolation,  
which exhibits a  similar static   phase transition,
$\mu$ is continuous and non-zero  at the threshold  \cite{Stauffer_92}. The situation is quite different in Axelrod's model as illustrated in 
the upper panel of Fig.\ \ref{fig:1}, which shows the mean density of domains as function of the Poisson parameter $q$.  The data suggest that for $q$ less than some critical value $q_c$ 
the density of domains vanishes in the thermodynamic limit and so that  there must exist a few macroscopic domains 
in that region.  For $q > q_c$ the number of domains  scales linearly with the number of sites in the lattice and so the average domain size 
$\langle S \rangle = L^2/ \mathcal{N} $ is finite in this region.  Since Fig.\ \ref{fig:1} indicates that the first derivative of $\mu$ is discontinuous at $q_c$ and that $\mu$ behaves as an order parameter of the  model, we will assume that $\mu \sim  \left ( q - {q}_c \right )^{\beta}$ near the critical point, where $\beta > 0$ is a critical exponent. In addition, for finite but large $L$ the finite size scaling theory yields \cite{Privman_90}
\begin{equation}\label{scal1}
\mu \sim L^{-\beta/\nu} f \left [ L^{1/\nu} \left ( q -  q_c \right ) \right ],
\end{equation}
where the scaling function is $f \left ( x \right ) \propto x^\beta$ for $x \gg 1$ and $\nu > 0$ is a critical exponent that determines the size of the critical region for finite $L$ and governs the divergence of the correlation length as $q \to q_c^+$.

\begin{figure}[!ht]
  \begin{center}
\includegraphics[width=0.48\textwidth]{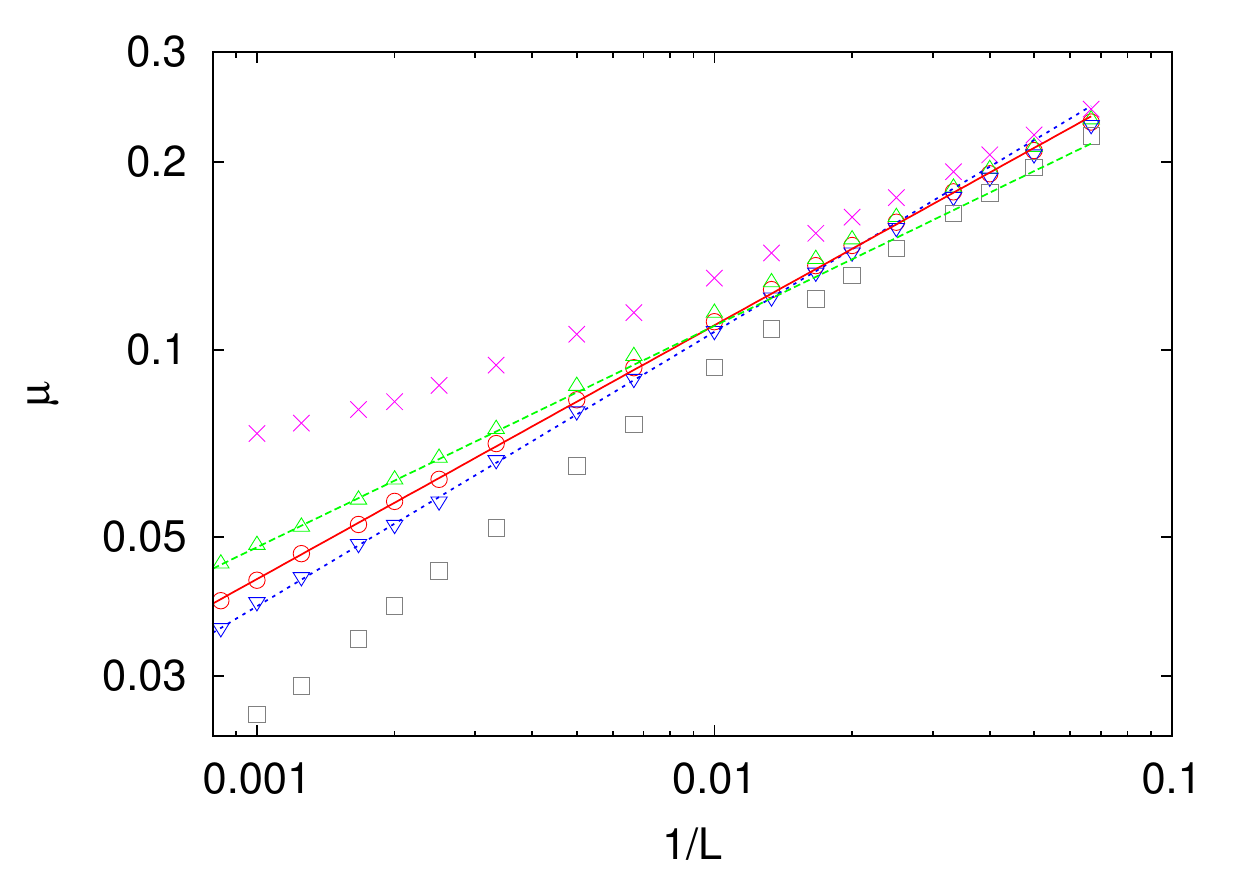}
  \end{center}
\caption{Log-log plot of the mean density of domains against the reciprocal of the linear lattice size for 
 $q =3.2$ ($\times$), $3.12$ ($\triangle$), $3.10$  ($\bigcirc$), $3.08$ ($\triangledown$), and $3.0$  ($\square$) . 
The error bars are smaller than the symbol sizes. The curve fitting the data for $q=3.10$ (solid line) is  $\mu = \mathcal{B} L^{\beta/\nu}$ with $\mathcal{B}=0.72 \pm 0.03$ and $\beta/\nu= 0.41 \pm 0.01$. }
\label{fig:2}
\end{figure}

Use of the finite size scaling equation (\ref{scal1}) allows us to produce quantitative estimates for the critical point $q_c$ and for the critical exponents $\beta$ and $\nu$, as well.   For instance,  according to that equation, $\mu$  should decrease
to zero as a power law of $L$ at $q= q_c$ and in Fig.\  \ref{fig:2} we explore this  fact to determine $q_c$ and the ratio $\beta/\nu$.
In particular, we fit the data for different values of $q$ with the function $\mu = \mathcal{B} L^{\beta/\nu}$ in the range $L \in \left [ 400, 1200 \right ]$ and gauge how the
fitting curves deviate from the data for $L \in \left [ 15, 300 \right ]$ in order to pick the critical value $q_c$. This is necessary because  for large $L$ 
all fittings are bona fide straight lines  in the log-log scale of  Fig.\  \ref{fig:2} in the range $q \in \left [ 0.308, 0.312 \right ]$. 
The data for $q = 3.12$ exhibits a definite convexity and the fitting with the exponent $ \left ( \beta/\nu \right )_{q=0.312} = 0.35 \pm 0.01$ deviates from the data already for  $L < 300$, whereas the data for $q = 3.08$ exhibits a very light concavity and the fitting with the exponent $ \left ( \beta/\nu \right )_{q=0.308} = 0.43 \pm 0.01$ deviates from the data only for $L < 50$. Finally, for $q = 0.310$ the fitting function  with the exponent $ \beta/\nu = 0.41 \pm 0.01$   fits the data very well  in the entire range of $L$ shown in the figure. Hence we conclude that
$q_c = 3.10 \pm 0.02$.

\begin{figure}[!ht]
  \begin{center}
\includegraphics[width=0.48\textwidth]{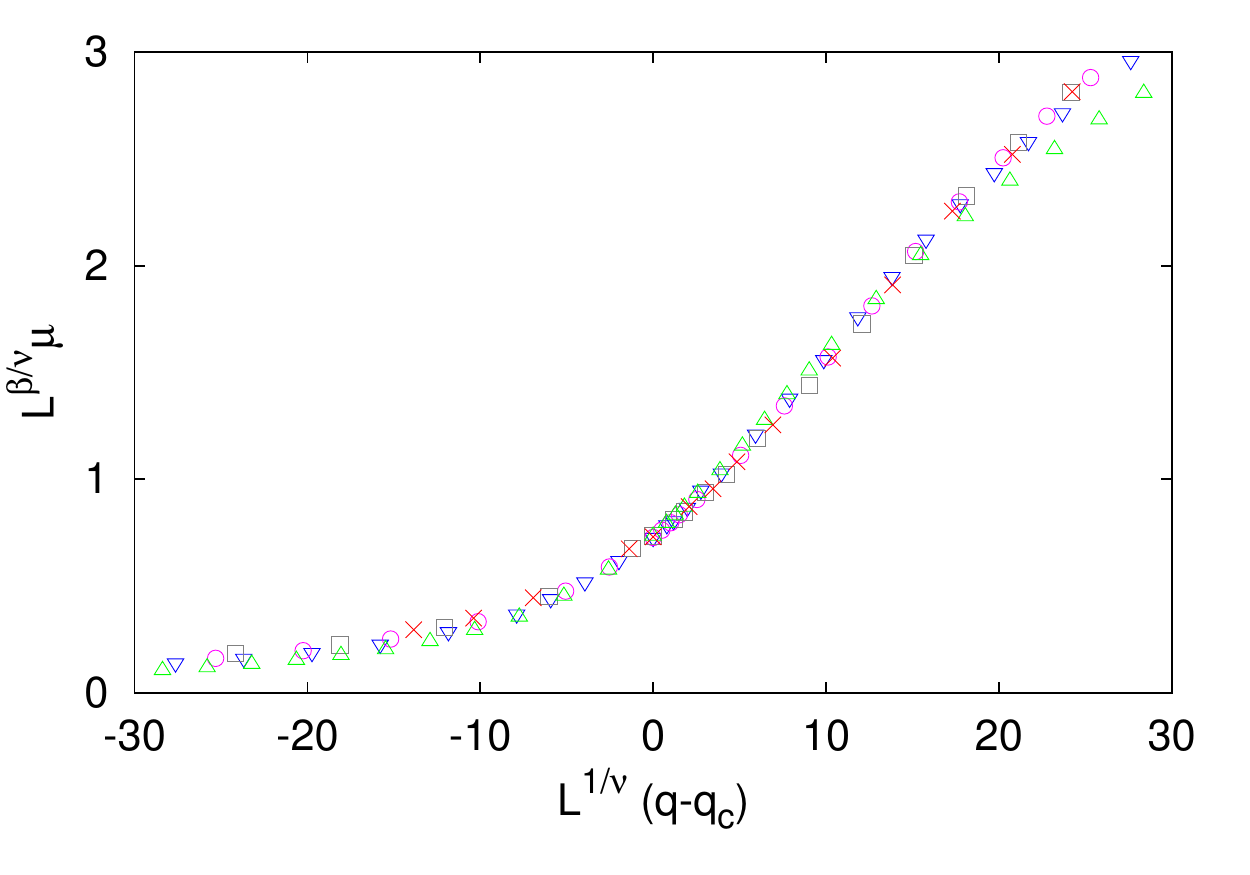}
  \end{center}
\caption{ Scaled mean density of domains against the scaled distance to the critical point 
for  lattices of linear  size  $L =  200 $ ($\triangle$),
$L=400$ ($\triangledown$), $L=600$ ($\bigcirc$) , $ L = 800$  ($\square$) and $L=1000$ ($\times$).
The error bars are smaller than the symbol sizes.
 The parameters are $q_c=3.10$, $\beta/\nu =0.41$ and $\nu = 1.63$. }
\label{fig:3}
\end{figure}

Once we have a good estimate for $q_c$ the best strategy is return to Fig.\ \ref{fig:1}
and fit  the data for $L=1000$  in the region near $q_c = 3.1$ using the fitting function $\mu = \mathcal{A} \left ( q  - q_c \right )^\beta$,  where 
$\mathcal{A}$  and $\beta$ are the two adjustable parameters of the fitting. This procedure yields $\beta =0.67 \pm 0.01$ for
the the order parameter critical exponent. The goodness of the resulting fitting is shown in the lower panel 
of Fig.\ \ref{fig:1}, which plots $\mu$ as function of the distance to the critical point $q-q_c$ in a log-log scale.

Finally, since $\beta=  0.67 \pm 0.01$ and   $\beta/\nu = 0.41 \pm 0.01 $ imply $\nu = 1.63 \pm 0.04$  we can validate our estimates
of  the critical quantities by checking whether the  scaled mean density of domains $ L^{\beta/\nu} \mu$ is independent of the lattice size $L$ when plotted against the  scaled distance to the critical parameter $L^{1/\nu} \left ( q -  q_c \right )$ as predicted by eq.\ (\ref{scal1}). This is shown in Fig.\ \ref{fig:3}  and
the quality of the resulting data collapse  confirms the soundness of  our estimate of the critical exponents.

\begin{figure}[!ht]
  \begin{center}
\includegraphics[width=0.48\textwidth]{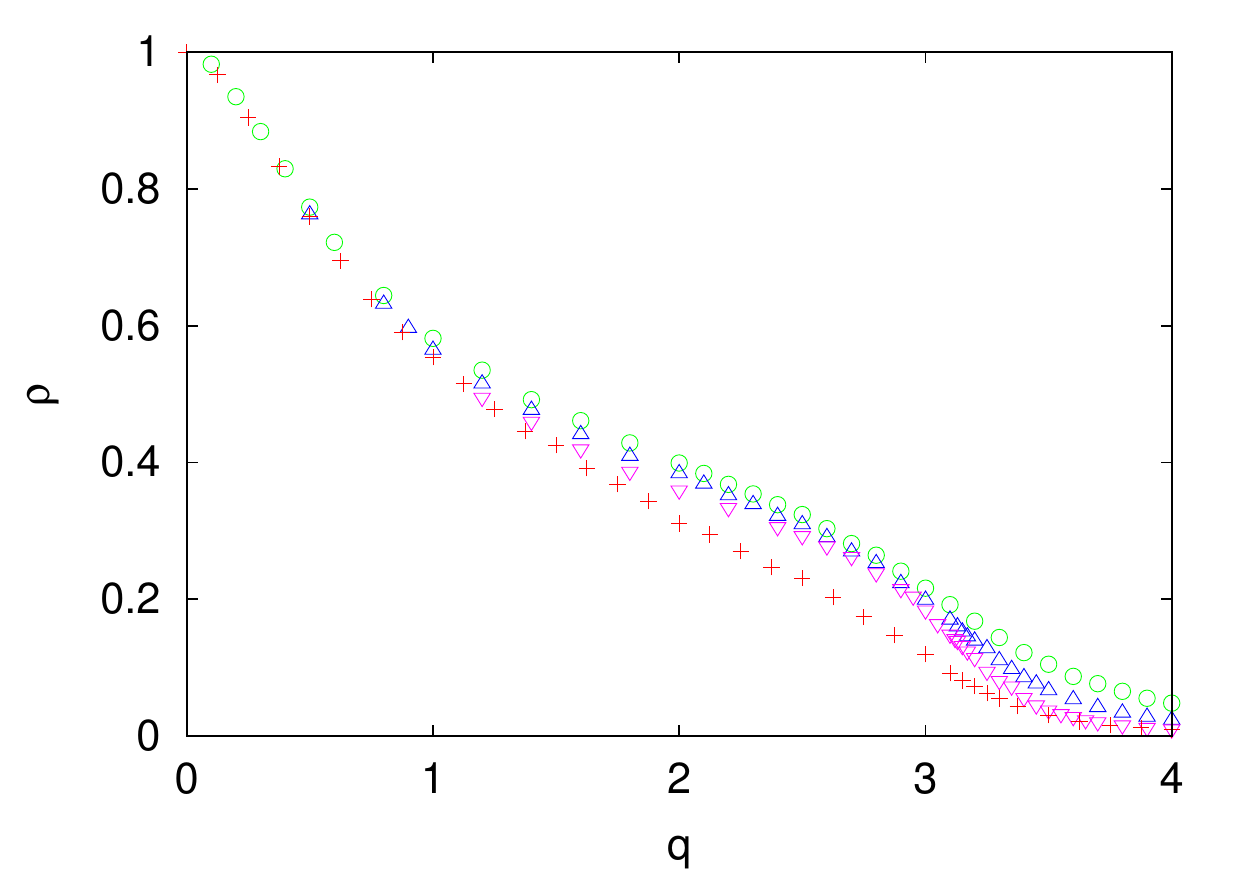}
  \end{center}
\caption{ Mean fraction of sites in the largest domain $\rho$  as function of the Poisson parameter $q$
for  lattices of linear  size $L = 100$  ($\bigcirc$), $L =  200 $ ($\triangle$) and 
$L=400$ ($\triangledown$).  The symbols $+$ show the results for a lattice with  $L=400$  and free boundary conditions.  The error bars are smaller than the symbol sizes.
 }
\label{fig:4}
\end{figure}

Let us  consider now  the standard order parameter  of Axelrod's model, namely, the mean fraction of lattice sites that belong to  the largest domain  $\rho = \left \langle \mathcal{S}_{max} \right \rangle/L^2$  (see, e.g., \cite{Castellano_00,Vilone_02,Klemm_03a}).  Figure \ref{fig:4} shows the dependence of $\rho$ on the Poisson parameter $q$. The finite size effects on $\rho$ are truly perplexing: for $q \to q_c^-$ the
results for  different  $L$ seem to  converge quickly  to some limiting value as $L$ increases but, surprisingly, as $q$ departs from $q_c$  in the region $q < q_c$ those results begin to diverge, and only for  small values of  $q$ (typically $q< 1$ for the lattice sizes
shown in the figure) we regain the independence on the lattice size, as expected. This means that for finite $L$ the measure $\rho$  exhibits a plateau separating the small $q$  region from the critical region.
It is interesting that this plateau is a finite size effect of the periodic boundary conditions since our simulations for free boundary conditions
(i.e., agents in the corners and in the sides of the lattice have two and three neighbors, respectively),
 also shown in Fig.\ \ref{fig:4}, exhibit a commoner approach to the critical regime. 
In particular, for free boundary conditions $\rho$ becomes independent of the lattice size already  for $q< 2.5$.   We observed the same effect of the boundary conditions in the one-dimensional lattice as well. However, for both boundary conditions a plot of $\rho$ against $1/L$ for fixed $q$ near $q_c$,  as shown in Fig.\ \ref{fig:2} for $\mu$,  shows a tendency of the data to level off at intermediate values of $L$  and then resume their decrease towards their limiting values as  $L$ becomes very large.

Therefore due to the somewhat pathological dependence of the standard order parameter $\rho$ on the
lattice size $L$ and on the Poisson parameter $q$, a study of the nature of the phase transition of Axelrod's model based on this parameter only would be practically impossible: it is no wonder that \cite{Castellano_00} refrained even from offering an estimate for $q_c$. In addition, since the dynamics  takes  a very  long  time to relax to absorbing configurations characterized by macroscopic  cultural domains
(see, e.g., \cite{Biral_15} for the quantification of this finding for the one-dimensional lattice), the  simulations are typically much slower in the region $q < q_c$  where $\rho$ is nonzero  than
in the region $q > q_c$ where $\mu$ is nonzero. Interestingly, for $q < q_c$ the simulations with free boundary conditions are way faster than  with periodic conditions,  perhaps because the translational invariance of the lattice is broken  in the former case.

\begin{figure}[!ht]
  \begin{center}
\includegraphics[width=0.48\textwidth]{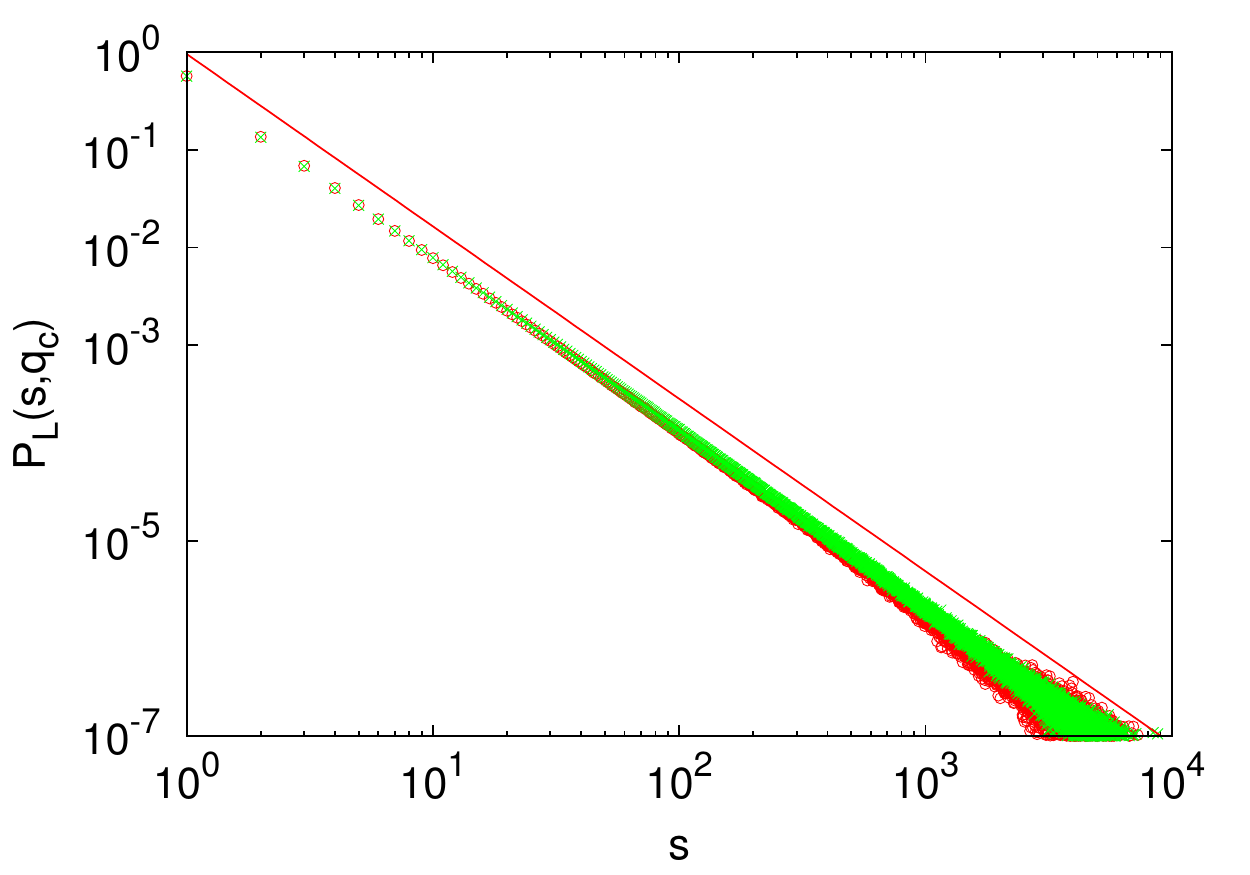}
  \end{center}
\caption{Probability distribution of the size $s$ of domains in a lattice of linear size $L$ at $q=q_c \approx 3.1$ for 
 $L = 100$  ($\bigcirc$), and $L =  200 $ ($\times$).  The line shows the power law $P_L \left ( s,q_c \right )  \sim s^{-\tau}$ 
 with  $\tau = 1.76$.    These distributions were generated using  $10^5$ independent runs for each $L$.
 }
\label{fig:5}
\end{figure}

Castellano et al.\ offered an   insight on the nature of the phase transition of Axelrod's model by focusing on the
probability distribution of domain sizes \cite{Castellano_00}. Consider the average domain size 
 \begin{equation}\label{S1}
 \langle S \rangle = \sum_{s=1}^{\infty} s P_L \left (s,q \right ) 
 \end{equation}
where $P_L \left (s,q \right ) $ is the probability distribution of the size $s $ of domains in a lattice of linear size $L$. Of course, 
$P_L \left (s,q \right )  = 0$ for $s > L^2$. In the limit $L \to \infty$ and for $q > q_c$  this probability can be written in the scaling form
$P_L \left (s,q \right ) = s^{-\tau} g \left ( s/s_{co}  \right )$ where $\tau > 0$ is the Fisher exponent and the scaling function
$g \left ( x \right )$ tends to a constant for $x \ll 1$ and decays very rapidly for $x \gg1$. As in the  standard percolation \cite{Stauffer_92},   
the transition occurs through the divergence of  the cutoff scale $s_{co} \sim \left ( q -q_c \right)^{-1/\sigma}  $ and hence of a correlation length $\xi$ since $s_{co} \sim \xi^{D} \sim \left ( q -q_c \right)^{-\nu D}$. Here $\sigma > 0$ is a critical exponent and $D \leq 2$ is the
fractal dimension of the incipient macroscopic domain. Clearly, $ \sigma = 1/\nu D$.
We note that the divergence of $ \langle S \rangle = L^2/ \mathcal{N}$  as  $q \to q_c^+$  implies that $\tau < 2$ \cite{Castellano_00} and
Fig.\ \ref{fig:5}, which shows the critical distribution $P_L \left (s,q_c \right )$, leads to the  estimate
$\tau = 1.76 \pm 0.01$. (The estimate of Castellano et al. is $\tau \approx 1.6$ for $L=100$.)

The finding that the density of domains  vanishes at $q =q_c$ as $\mu \sim \left ( q - q_c \right)^\beta$ makes the continuous transition of Axelrod's model to depart markedly from the percolation transition, since the exponent $\beta$ has no counterpart
in that case. In fact, we need to derive  the relations between $\beta$ and the other critical exponents anew. In particular, noting that
$  \langle S \rangle = 1/\mu$  we can use eq.\ (\ref{S1}) to obtain the relation $\beta = \left (2 - \tau \right )/\sigma$,  which together with
$\sigma = 1/\nu D$  result in the estimates $\sigma = 0.36 \pm 0.02 $ and $D = 1.71 \pm 0.02 $. 

We note that not  only our order parameter  ($\mu$) is different from the order parameter ($\rho$) considered 
in the original analysis of the continuous phase transition of Axelrod's  model \cite{Castellano_00} but  the regimes
investigated differ as well. In particular,  here we focus on the regime  $q>q_c$,
where $\mu >0$ and $\rho = 0$ in the thermodynamic limit,  and thus we describe the onset of order 
by focusing on the process of  agglutination of the  domains. Although this is 
different from observing the growing of a macroscopic domain in the regime $q < q_c$  as done by Castellano et al.\
\cite{Castellano_00}, both perspectives  describe the same critical phenomenon.

Our aim here was to offer a quantitative characterization of the continuous nonequilibrium phase transition  of the Poisson variant of Axelrod's  model that was first  reported in 2000 \cite{Castellano_00}. The transition is static in nature and separates two types of absorbing configurations that differ on their distributions of
domain sizes. Because of the  two distinctive features --  both phases correspond to  absorbing configurations and the density of domains vanishes at the critical point -- 
the continuous phase transition of Axelrod's model is characterized by a set of critical exponents that sets it apart from the known universality classes  of  nonequilibrium  lattice models \cite{Marro_99}.

\acknowledgments

This research was partially supported by grant
2013/17131-0, S\~ao Paulo Research Foundation
(FAPESP) and by grant 303979/2013-5, Conselho Nacional de Desenvolvimento 
Cient\'{\i}\-fi\-co e Tecnol\'ogico (CNPq). The research used resources of the LCCA - Laboratory of Advanced Scientific Computation of the University of S\~ao Paulo.


\begin{thebibliography}{99}

\bibitem{Lazarsfeld_48}  P. Lazarsfeld, B. Berelson  and H. Gaudet,
 {\it The People’s Choice}
 (Columbia University Press, New York, 1948).

\bibitem{Castellano_09} C. Castellano, S. Fortunato and V. Loreto,
{\it Rev. Mod. Phys.} {\bf 81}, 591 (2009).

\bibitem{Axelrod_97} R. Axelrod,  {\it J. Conflict Res.} {\bf  41}, 203 (1997).

\bibitem{Klemm_03a} K. Klemm, V. M. Egu\'{\i}luz, R. Toral and M. San Miguel, {\it Phys. Rev. E} {\bf 67}, 026120 (2003).

\bibitem{Castellano_00} C.  Castellano, M.  Marsili and A. Vespignani,
{\it Phys. Rev. Lett.} {\bf  85}, 3536 (2000).

\bibitem{Klemm_03} K. Klemm, V. M. Egu\'{\i}luz, R. Toral and M. San Miguel,  {\it Physica A} {\bf 327}, 1 (2003).

\bibitem{Vazquez_07} F. Vazquez and S. Redner, {\it Europhys. Lett.} {\bf 78}, 18002 (2007).

\bibitem{Barbosa_09} L.A.  Barbosa   and  J. F.  Fontanari,  {\it Theor. Biosci.} {\bf 128}  205 (2009).

\bibitem{Marro_99} J. Marro and R. Dickman, {\it Nonequilibrium Phase Transitions in 
Lattice Models} (Cambridge University Press, Cambridge, UK, 1999).

\bibitem{Jensen_93} I. Jensen and R. Dickman, {\it Phys. Rev. E} {\bf 48},  1710 (1993)

\bibitem{Vilone_02} D. Vilone, A. Vespignani and C. Castellano, {\it Europ. Phys. J. B} {\bf 30}, 399 (2002).

\bibitem{Stauffer_92}   D. Stauffer and A. Aharony, {\it Introduction to Percolation Theory}
 (Taylor \& Francis, London, 1992).

\bibitem{Privman_90} V. Privman, {\it Finite-Size Scaling and Numerical Simulations of Statistical Systems}
(World Scientific, Singapore, 1990).

\bibitem{Biral_15}
E.J.P. Biral, P. F. C. Tilles and J. F. Fontanari, {\it J. Stat. Mech. } P04006 (2015).




\end{thebibliography}
\end{document}